\documentclass[journal]{IEEEtran}
\usepackage{amsmath,amsfonts}
\usepackage{algorithmic}
\usepackage{algorithm}
\usepackage{array}
\usepackage[caption=false,font=normalsize,labelfont=sf,textfont=sf]{subfig}
\usepackage{textcomp}
\usepackage{stfloats}
\usepackage{url}
\usepackage{verbatim}
\usepackage{graphicx}
\usepackage{cite}
\usepackage{mathtools}
\usepackage{threeparttable}
\usepackage{multirow}
\usepackage{siunitx}
\usepackage{booktabs}

\usepackage[colorlinks, linkcolor=blue, anchorcolor=black, citecolor=blue]{hyperref}
\begin{document}
\captionsetup[figure]{name={Fig.},labelsep=period,singlelinecheck=off}

\title{CoDS: Collaborative Perception via Digital Semantic Communication}

\author{

    Jipeng~Gan,~\IEEEmembership{Student Member,~IEEE,}  
    Le~Liang,~\IEEEmembership{Member,~IEEE,}
    Hua~Zhang,~\IEEEmembership{Member,~IEEE,}
    
    Chongtao Guo,~\IEEEmembership{Member,~IEEE,}
    and~Shi~Jin,~\IEEEmembership{Fellow,~IEEE}

  \thanks{Jipeng~Gan, Le Liang, Hua~Zhang and Shi Jin are with the National Mobile Communications Research Laboratory, Southeast University, Nanjing 210096, China (e-mail: \{jpgan, lliang, huazhang, jinshi\}@seu.edu.cn). Le Liang is also with the Purple Mountain Laboratories, Nanjing 211111, China.

  Chongtao Guo is with the College of Electronics and Information Engineering, Shenzhen University, Shenzhen 518060, China (e-mail: ctguo@szu.edu.cn).
}
  }

\maketitle

\begin{abstract} 
Semantic communication has been introduced into collaborative perception systems for autonomous driving, offering a promising approach to enhancing data transmission efficiency and robustness. Despite its potential, existing semantic communication approaches predominantly rely on analog transmission models, rendering these systems fundamentally incompatible with the digital architecture of modern vehicle-to-everything (V2X) networks and posing a significant barrier to real-world deployment.
To bridge this critical gap, we propose CoDS, a novel collaborative perception framework based on digital semantic communication, designed to realize semantic-level transmission efficiency within practical digital communication systems. Specifically, we develop a semantic compression codec that extracts and compresses task-oriented semantic features while preserving downstream perception accuracy. Building on this, we propose a novel semantic analog-to-digital converter that converts these continuous semantic features into a discrete bitstream, ensuring integration with existing digital communication pipelines. 
\textcolor{black}{Furthermore, we develop an uncertainty-aware network (UAN) that assesses the reliability of each received feature and discards those corrupted by decoding failures, thereby mitigating the cliff effect of conventional channel coding schemes under low signal-to-noise ratio (SNR) conditions.}
Extensive experiments demonstrate that CoDS significantly outperforms existing semantic communication and traditional digital communication schemes, achieving state-of-the-art perception performance while ensuring compatibility with practical digital V2X systems.
\end{abstract} 


\section{Introduction}	
\begin{figure}[t!]
  \centering
  \includegraphics[width=3.2in]{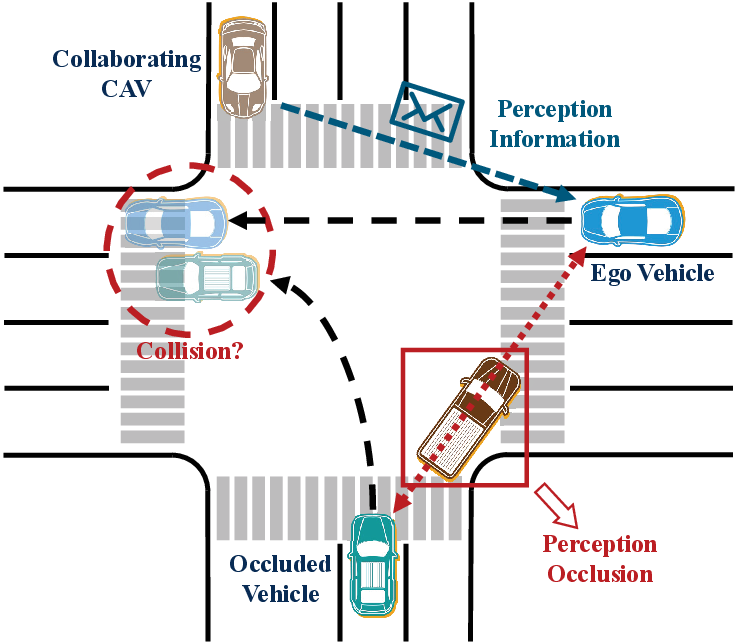}
  \caption{An illustrative example of collaborative perception scenario.}
  \label{fig1}
\end{figure}

Autonomous driving is a foundational component of modern intelligent transportation systems, and its success critically relies on accurate and robust environmental perception  \cite{av, av2, AV3}. This perception capability enables essential downstream tasks such as object detection \cite{objectdetection1}, tracking \cite{tracking}, and motion prediction \cite{motionprediction}, which serve as the foundation for safe and efficient vehicle planning and control \cite{downstream}. 
Conventionally, vehicles achieve environmental awareness through a stand-alone perception approach, relying exclusively on their own on-board sensors like cameras, millimeter-wave radar, and particularly light detection and ranging (LiDAR) devices, which provide precise 3D spatial information \cite{sensors, lidar}.
However, these systems are constrained by inherent limitations, including a restricted field of view and critical blind spots caused by occlusions from buildings or other vehicles, which may compromise safety. 
To overcome these challenges, collaborative perception has emerged as a promising paradigm. By leveraging vehicle-to-everything (V2X) communication, connected automated vehicles (CAVs) can share and fuse sensory data from multiple viewpoints to generate comprehensive scene representations \cite{CP1, CP2, cp3 }. 
As shown in Fig.~\ref{fig1}, sharing perception data from a collaborating CAV allows the ego vehicle to overcome line-of-sight occlusions. This enables the detection of hidden objects and planning of safe trajectories, directly enhancing its safety and robustness.

Currently, three primary strategies have been developed to implement collaborative perception: early, intermediate, and late collaboration, each with distinct communication and computational resource requirements \cite{fusion}. Specifically, early collaboration involves sharing raw sensor data, such as LiDAR point clouds, which preserves complete scene information and thus offers the highest potential perception fidelity \cite{early1}. However, its prohibitively high bandwidth consumption may render it impractical for real-time applications. Conversely, late collaboration minimizes communication overhead by transmitting only final object-level detections \cite{late1}. However, such efficiency comes at the cost of discarding rich contextual information, which limits perception performance in complex scenarios. As a balanced solution, intermediate collaboration compresses raw sensor data into compact feature representations for transmission \cite{inter1}. This approach significantly reduces transmission overhead compared to early collaboration while maintaining comparable perception performance, thus achieving an effective balance between accuracy and communication efficiency. Consequently, intermediate collaboration has become a dominant paradigm for communication in resource-constrained collaborative perception systems.

Within the intermediate collaboration paradigm, significant research has focused on optimizing the balance between perception performance and communication overhead \cite{interfusion}. For instance, several studies have concentrated on intelligent information selection schemes to determine what, when, and to whom to communicate \cite{when2comm,how2comm,where2comm,v2vnet}, as well as MAC-layer optimization for efficient resource allocation to ensure reliable V2X communication \cite{mac1,mac2,mac3}. Despite these advances, a critical limitation of most existing collaborative perception frameworks is the implicit assumption of an ideal, error-free communication channel. This critical oversight divorces the perception schemes from the physical reality of wireless transmission, where inevitable data corruption can render a CAV's perception dangerously incomplete or incorrect, directly compromising safety.

To address this limitation, recent research has begun to integrate semantic communication into collaborative perception frameworks \cite{franklin,scomcp,semantic1,harq, semantic3}. The dominant approach in these pioneering works is analog semantic communication (ASC), which is based on an end-to-end joint source-channel coding methodology. By directly mapping perception features to channel symbols, these methods demonstrated a profound potential for optimizing the communication system directly towards the final perception performance (e.g., object detection precision), rather than for traditional metrics like bit-level fidelity. For instance, the seminal work in \cite{franklin} introduced a task-oriented semantic codec, demonstrating transmission efficiency and resilience to noise under low signal-to-noise ratio (SNR) conditions. Building on this, subsequent efforts have enhanced efficiency and robustness through more advanced techniques, such as Transformer-based semantic codecs \cite{scomcp}, masked autoencoder pre-training \cite{semantic1}, and semantic-aware hybrid automatic repeat request (HARQ) protocols \cite{harq}. Collectively, these advances have established semantic communication as a new paradigm capable of achieving high bandwidth efficiency and perception performance. 
 
While ASC has demonstrated significant potential, its underlying design philosophy creates a critical implementation gap with the standardized, modular architecture of real-world vehicular networks. 
Modern systems like cellular-V2X (C-V2X) are built on a robust digital pipeline comprising discrete processes for channel coding (e.g., low-density parity-check (LDPC) codes), modulation, etc.
\textcolor{black}{By circumventing this infrastructure, the end-to-end analog design of the ASC can achieve optimal theoretical performance. However, this departure from the digital pipeline presents significant deployment challenges and results in the forfeiture of mature, robust error correction capabilities inherent in standardized channel coding schemes.}
To bridge the gap between theoretical potential and practical deployment, digital semantic communication has emerged as a compelling solution \cite{dsc1, dsc2}. Unlike its analog counterpart, digital semantic communication discretizes semantic information into compact bitstreams. This approach retains the core benefit of semantic-level transmission efficiency while ensuring full compatibility with existing digital communication pipelines. Consequently, digital semantic communication can leverage the powerful error-correction capabilities of established channel codes and integrate with MAC-layer strategies for scalable multi-agent resource management. Thus, digital semantic communication provides a practical and robust pathway toward deploying intelligent, efficient, and scalable V2X collaborative perception systems.

In response, this paper proposes a novel intermediate collaborative perception framework based on digital semantic communication (CoDS). To the best of our knowledge, this work represents the first semantic communication system specifically designed for compatibility with digital communication networks for collaborative perception.
Particularly, we first develop a semantic compression codec that extracts key semantic features to reduce the bandwidth requirements. Building on this, we propose a semantic analog-to-digital converter that maps continuous-valued semantic features into discrete bitstreams, thereby enabling integration with existing digital communication systems. Furthermore, we develop an uncertainty-aware network (UAN) to mitigate the performance degradation of conventional channel coding schemes under low SNR conditions. The main contributions of this work are summarized as follows:
\begin{itemize}
\item We propose a digital semantic communication framework for collaborative perception. This framework is designed to be fully compatible with modern digital communication architectures and enables efficient semantic-level information sharing between collaborative CAVs.
\item We develop a semantic compression codec, constructed from cascaded down/up-sampling blocks, that is end-to-end optimized for the perception task. The codec learns to extract and compress only the most task-oriented semantic features, achieving significant bandwidth reduction while preserving downstream perception accuracy.
\item We design a semantic analog-to-digital converter for bidirectional conversion between continuous semantic features and discrete bitstreams. 
This converter utilizes a decoupling layer to project features into independent subspaces, which are then quantized by mapping to the nearest entry in corresponding semantic embedding spaces and encoded as discrete bitstreams.
\item We develop a UAN that leverages soft information from the channel decoder and semantic similarity between semantic features to assess the post-transmission reliability of received semantic features. This allows the system to selectively retain or discard received semantic features, mitigating the impact of performance collapse in low SNR environments.
\end{itemize}

The remainder of this paper is organized as follows. Section II presents the system model of the proposed CoDS framework. Section III details the proposed semantic compression codec, semantic analog-to-digital converter, and UAN, along with the corresponding loss function design and training strategies. Section IV provides simulation results to demonstrate the advantages of the proposed framework. Finally, Section V concludes the paper.

\section{System Model}

\begin{figure*}[htbp]
  \centering
  \includegraphics[width=1\linewidth]{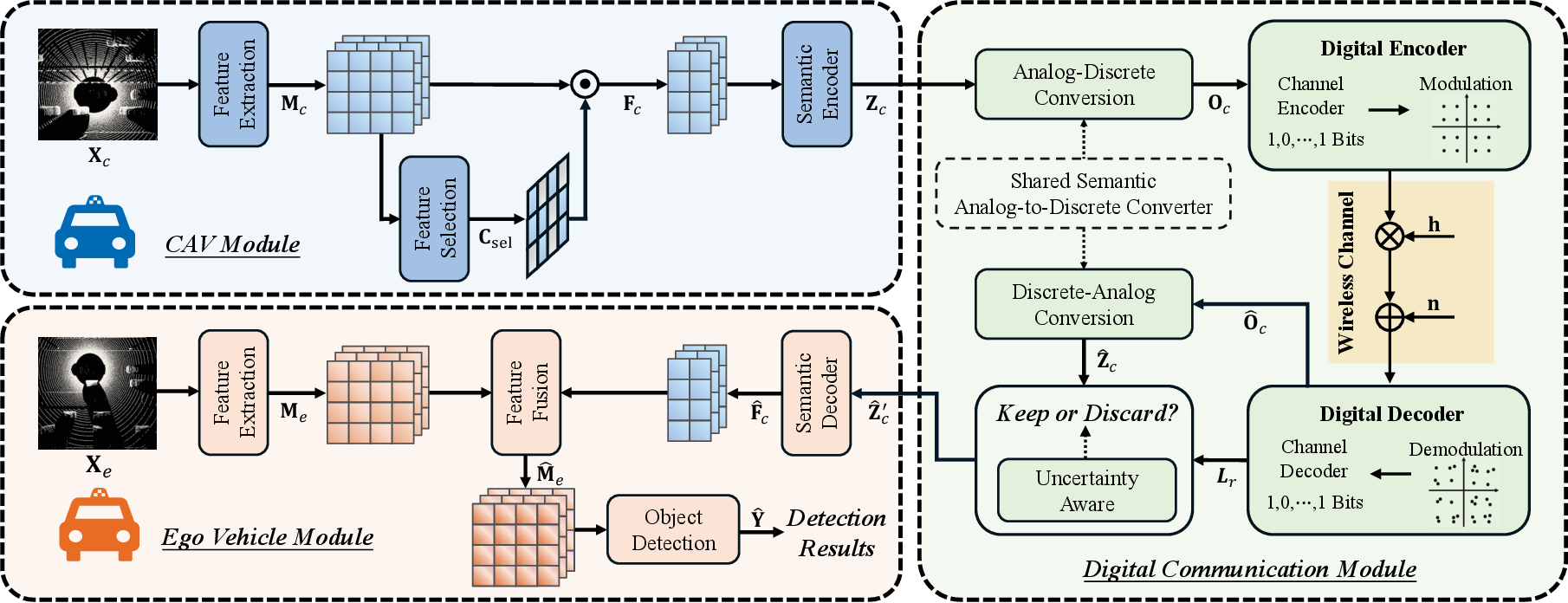}
  \caption{An illustration of the proposed CoDS framework.}
  \label{fig:system_model}
\end{figure*}

This section details the proposed collaborative perception framework based on digital semantic communication. As illustrated in Fig.~\ref{fig:system_model}, we consider a representative vehicle-to-vehicle (V2V) scenario in which a single collaborating CAV transmits its LiDAR-based perception information to an ego vehicle to enable intermediate collaboration. The overall architecture comprises three primary components: the CAV module, the ego vehicle module, and the digital communication module. Specifically, the CAV module processes its raw LiDAR point cloud and encodes it into a compact semantic feature representation for transmission. Subsequently, the ego vehicle module fuses the received semantic features with its own local sensor data to perform a 3D object detection task. Meanwhile, the digital communication module is responsible for transmitting semantic features from the CAV to the ego vehicle over a wireless channel. This process involves quantization, channel coding, and modulation at the transmitter, followed by demodulation, decoding, and reliability assessment at the receiver.
 
\subsection{CAV Module}

Raw LiDAR point clouds are usually unstructured and sparse, making them unsuitable for direct processing by downstream perception tasks. To address this, the CAV first employs a feature extraction network to transform its raw LiDAR point cloud, $\mathbf{X}_c$, into a bird's-eye view (BEV) feature map, expressed as
\begin{equation}
\label{eq:bev_extraction}
\mathbf{M}_c = \mathcal{F}_{\alpha}(\mathbf{X}_c),
\end{equation}
where $\mathbf{M}_c \in \mathbb{R}^{H \times W \times C}$ denotes the BEV feature map, with spatial dimensions $H \times W$ and $C$ feature channels.  $\mathcal{F}_{\alpha}(\cdot)$ represents the feature extraction network, parameterized by $\alpha$, which is adapted from the backbone module of PointPillars~\cite{pointpillars}.

Given that the CAV's feature map, $\mathbf{M}_c$, is typically sparse in the spatial dimension, transmitting the full feature map is inefficient. Therefore, we employ a feature selection network $\mathcal{F}_{\beta}(\cdot)$, parameterized by $\beta$, to identify important perception features for transmission, thereby reducing communication overhead \cite{where2comm}. Specifically, the network $\mathcal{F}_{\beta}(\cdot)$ processes the CAV's feature map $\mathbf{M}_c$ to generate a binary selection mask $\mathbf{C}_{\text{sel}} \in \{0, 1\}^{H \times W}$, where a value of 1 indicates a spatial location selected for transmission and 0 indicates that the location is discarded. This operation is formulated as
\begin{equation} 
\label{eq3} 
\mathbf{C}_{\text{sel}} = \mathcal{F}_{\beta}(\mathbf{M}_c). 
\end{equation} 

The mask, $\mathbf{C}_{\text{sel}}$, is then broadcast along the channel dimension and applied to the feature map via element-wise multiplication. The remaining feature vectors are gathered into a feature matrix, $\mathbf{F}_c \in \mathbb{R}^{K \times C}$, which contains only the selected features. Here, $K$ denotes the number of selected features, equal to the $\ell_1$ norm, $||\cdot||_1$, of the mask, i.e., $K = ||\mathbf{C}_{\text{sel}}||_1$. This selection and collection process is abstractly represented as
\begin{equation} 
\label{eq:feature_select} 
\mathbf{F}_c = \text{Gather}(\mathbf{M}_c, \mathbf{C}_{\text{sel}}), 
\end{equation} 
where $\text{Gather}(\cdot, \cdot)$ represents the operation of masking the feature map and collecting the selected features.
Moreover, we define the spatial compression ratio as the fraction of features selected for transmission,  given by $\gamma_s = (H \times W)/ K$.

To further mitigate communication overhead along the channel dimension, the CAV employs a semantic compression encoder $\mathcal{F}_{\theta}(\cdot)$, parameterized by $\theta$, to extract key semantic information from the selected feature matrix $\mathbf{F}_c$, thereby compressing it into a compact semantic feature representation, given by
\begin{equation} 
\label{eq:semantic_encoding} 
\mathbf{Z}_c = \mathcal{F}_{\theta}(\mathbf{F}_c),
\end{equation}
where $\mathbf{Z}_c \in \mathbb{R}^{K \times C'}$ represents the semantic features, and $C'$ denotes the number of channels after compression, with $C' < C$. The channel compression ratio is defined as the ratio of the number of input to output channels, i.e., $\gamma_c = {C}/{C}'$.

\subsection{Ego Vehicle Module}

At the ego vehicle side, the received semantic features, denoted as $\mathbf{\hat{Z}}'_c$, are decompressed back to the original channel dimension using a semantic compression decoder $\mathcal{F}_{\phi}(\cdot)$, parameterized by $\phi$. This decompression process is expressed as
\begin{equation}
    \label{eq:semantic_decoding}
    \mathbf{\hat{F}}_c = \mathcal{F}_{\phi}(\mathbf{\hat{Z}}'_c),
\end{equation}
where $\mathbf{\hat{F}}_c$ represents the decompressed CAV feature matrix.
Then, the ego vehicle fuses its local feature map $\mathbf{M}_e$ with the reconstructed features $\mathbf{\hat{F}}_c$ using an attention-based fusion function $f_{\mu}(\cdot)$  \cite{opencood}. This fusion process aggregates information from both the ego and CAV to yield a more comprehensive scene representation, written as
\begin{equation}
    \label{eq:feature_fusion}
    \mathbf{\hat{M}}_e = f_{\mu}(\mathbf{M}_e, \mathbf{\hat{F}}_c),
\end{equation}
where $\mathbf{\hat{M}}_e$ represents the fused feature map.
Finally, the fused feature map $\mathbf{\hat{M}}_e$ is processed by a detection network, $\mathcal{F}_{\kappa}(\cdot)$, to yield the final perception result, $\mathbf{\hat{Y}}$. The detection process is given by
\begin{equation}
    \label{eq:object_detection}
    \mathbf{\hat{Y}} = \mathcal{F}_{\kappa}(\mathbf{\hat{M}}_e).
\end{equation}

The network $\mathcal{F}_{\kappa}(\cdot)$, parameterized by $\kappa$, is composed of two parallel heads: a regression head and a classification head. The regression head estimates the seven parameters of each 3D bounding box, including its center $(x, y, z)$, dimensions $(w, l, h)$, and yaw angle $\theta$.  Simultaneously, the classification head determines a confidence score for each predicted bounding box, representing its object category likelihood.

\subsection{Digital Communication Module}

To facilitate the transmission of semantic features from the CAV to the ego vehicle, a digital communication module is designed, consisting of a transmitter, a receiver, and a wireless channel.

At the transmitter, for compatibility with the digital communication framework, the continuous-valued semantic features $\mathbf{Z}_c$ need to be converted into discrete bitstreams. This is accomplished using a semantic analog-to-digital converter $\mathcal{F}_{\omega}(\cdot)$, parameterized by $\omega$, expressed as
\begin{equation}
    \label{eq:semantic_quantization}
    \mathbf{O}_c = \mathcal{F}_{\omega}(\mathbf{Z}_c),
\end{equation}
where $\mathbf{O}_c \in \{0,1\}^{K \times q}$ is the discrete bitstream and $q$ denotes the number of quantization bits for each semantic feature.
The bitstream is then processed by a digital encoder that applies channel coding and quadrature amplitude modulation (QAM) to produce a sequence of complex-valued channel symbols, $\mathbf{S}_c \in \mathbb{C}^{D}$, where $D$ denotes the number of channel symbols.

After encoding, the symbols $\mathbf{S}_c$ are transmitted to the ego vehicle via a noisy physical channel. The channel model is formulated as
\begin{equation}
    \mathbf{\hat{S}}_c = \mathbf{h} \odot \mathbf{S}_c + \mathbf{n},
\end{equation}
where $\odot$ denotes element-wise multiplication, $\mathbf{\hat{S}}_c \in \mathbb{C}^{D}$ denotes the received symbols, $\mathbf{h} \in \mathbb{C}^{D }$ represents the channel gain coefficients, and $\mathbf{n} \in \mathbb{C}^{D }$ is the additive white Gaussian noise (AWGN). The noise $\mathbf{n}$ is assumed to consist of independent and identically distributed (i.i.d.) samples following a circularly symmetric complex Gaussian distribution, i.e., $\mathbf{n} \sim \mathcal{CN}(0, \sigma^2 \mathbf{I})$, where $\sigma^2$ is the noise variance and $\mathbf{I}$ is the identity matrix.

At the receiver, the received symbols $\mathbf{\hat{S}}_c$ are processed by a digital decoder to reconstruct the original bitstream. The digital decoder performs both demodulation and channel decoding, producing a hard-decision bitstream $\mathbf{\hat{O}}_c \in \{0,1\}^{K \times q}$ and the corresponding soft-decision information $\mathbf{L}_r \in \mathbb{R}^{K \times q}$, which represents the log-likelihood ratio (LLR) matrix.
Subsequently, the received bitstream $\mathbf{\hat{O}}_c$ is reconstructed into semantic features, $\mathbf{\hat{Z}}_c \in \mathbb{R}^{K \times C'}$, through the inverse semantic analog-to-digital converter operation, described as
\begin{equation}
    \label{eq:semantic_dequantization}
    \mathbf{\hat{Z}}_c = \mathcal{F}_{\omega}^{-1}(\mathbf{\hat{O}}_c),
\end{equation}
where $\mathcal{F}_{\omega}^{-1}(\cdot)$ denotes the inverse function corresponding to the semantic analog-to-digital converter, $\mathcal{F}_{\omega}(\cdot)$.

However, the reconstructed semantic features, $\mathbf{\hat{Z}}_c$, may contain errors due to channel noise, resulting in a degraded perception quality. To mitigate this, the ego vehicle employs a UAN, $\mathcal{F}_{\varphi}(\cdot)$ parameterized by $\varphi$, to assess the reliability of the received semantic features. Specifically, the network processes the LLR matrix, $\mathbf{L}_r$, and outputs confidence scores, $\mathbf{p} \in [0, 1]^K$, for each semantic feature, as follows
\begin{equation}
    \label{eq:gating_prob}
    \mathbf{p} = \mathcal{F}_{\varphi}(\mathbf{L}_r),
\end{equation}
where each element in $\mathbf{p}$ represents the confidence that the corresponding semantic feature has been decoded correctly.
Subsequently, the ego vehicle employs a gating mechanism to selectively filter the received semantic features based on their confidence scores, generating a binary gating mask, $\mathbf{C}_{\text{unc}} \in \{0, 1\}^K$. Each semantic feature is retained if its confidence score exceeds a predefined threshold, $\tau_{\text{the}}$, indicating higher reliability in the decoded information. The gating operation is formulated as 
\begin{equation}
    \label{eq:gating_mask}
    \mathbf{C}_{\text{unc}} = \mathbf{I}(\mathbf{p} > \tau_{\text{the}}),
\end{equation}
where $\mathbf{I}(\cdot)$ denotes the element-wise indicator function that outputs 1 for the entries with the corresponding condition satisfied, and 0 otherwise.
The binary gating mask, $\mathbf{C}_{\text{unc}}$, is then applied to filter unreliable components out of the reconstructed semantic features, $\mathbf{\hat{Z}}_c$. This is achieved by broadcasting the mask along the channel dimension and applying element-wise multiplication, which yields the filtered semantic feature representation as follows
\begin{equation}
    \label{eq:gating_application}
    \mathbf{\hat{Z}}'_c =\text{Gather}(\mathbf{\hat{Z}}_c, \mathbf{C}_{{\text{unc}}}),
\end{equation} 
where $\mathbf{\hat{Z}}'_c \in \mathbb{R}^{K' \times C'}$ contains only high-confidence semantic features, with $K'$ denoting the number of retained semantic features.

\section{Proposed CoDS Framework}

In this section, we first introduce the semantic compression codec and the semantic analog-to-digital converter, followed by their loss functions and the corresponding training strategy. The UAN and its associated training algorithm are then described in detail.

\begin{figure}[t!]
  
  \centering
  \includegraphics[width=3.5in]{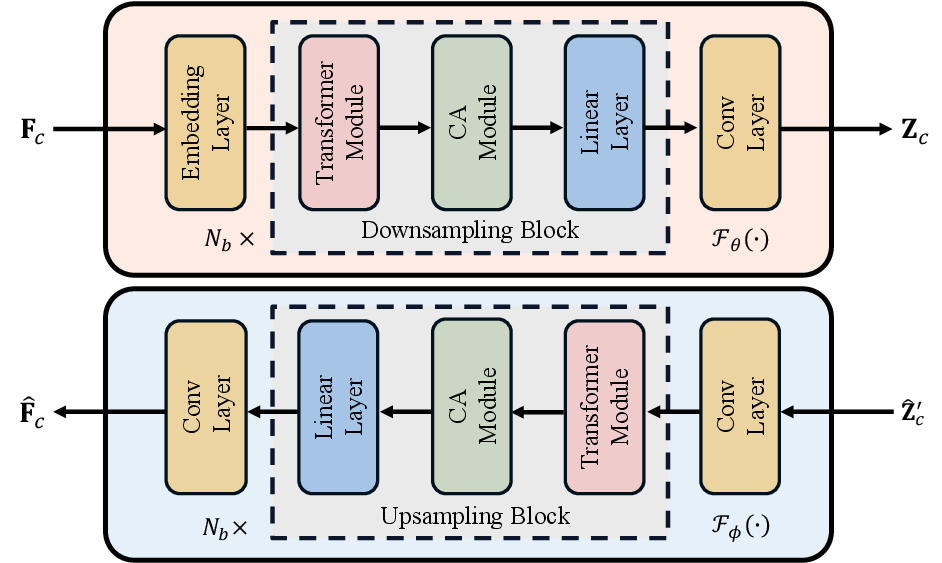}
  \caption{Network structure of our proposed semantic compression codec.}
  \label{fig:semantic_codec}
\end{figure}

\subsection{Semantic Compression Codec}
This subsection introduces the proposed semantic compression codec, denoted as $\mathcal{F}_{\theta}(\cdot)$ in (4) and $\mathcal{F}_{\phi}(\cdot)$ in (5),  with its architecture illustrated in Fig.~\ref{fig:semantic_codec}. The feature matrix, $\mathbf{F}_c$, contains significant redundancy along the channel dimension, leading to excessive communication overhead. To this end, we propose a semantic compression codec that extracts perception-relevant information into compact semantic feature representations, thereby reducing transmission overhead while preserving perception performance. The codec comprises a semantic compression encoder at the CAV and a corresponding semantic compression decoder at the ego vehicle, which work in tandem to compress and reconstruct the features.

At the CAV, the semantic compression encoder is primarily composed of a stack of $N_b$ downsampling blocks, where each block achieves a two-fold channel compression ratio. Consequently, the number of blocks is determined by $N_{b} = \log_2(\gamma_c)$. Each block consists of three key components: a Transformer module, a channel attention (CA) module, and a linear layer~\cite{scomcp}.
This hierarchical architecture enables the encoder to progressively extract and compress the most salient features through multiple stages of refinement. Specifically, within each block, the Transformer module captures global contextual information by modeling long-range dependencies within the features. The CA module then processes the Transformer output, dynamically computing channel-wise scaling and offset parameters to selectively enhance the most informative channels. Finally, the linear layer performs dimensionality reduction by compressing the channel dimension of the attention-refined features, thereby achieving the desired level of compression.
 
Among these components, the CA module first aggregates spatial information through parallel average-pooling and max-pooling operations applied to the input features, $\mathbf{f}$. The resulting pooled features are processed by linear layers and concatenated to form an aggregated spatial context vector, $\mathbf{w}$. This context vector is then fed into two independent linear layers to generate a channel-wise scaling vector, $\text{Scale}(\mathbf{w})$, and an offset vector, $\text{Offset}(\mathbf{w})$. The module applies these vectors in a feature-specific affine transformation, formulated as
\begin{equation}
    \label{eq:ca_affine_transform}
    \mathbf{f}' = \text{Scale}(\mathbf{w}) \odot \mathbf{f} + \text{Offset}(\mathbf{w}),
\end{equation}
where $\mathbf{f}'$ represents the attention-refined features. 
At the ego vehicle, the semantic compression decoder mirrors the encoder's architecture, employing the same number of upsampling blocks for symmetric reconstruction. This design allows the decoder to progressively restore the features from their compressed representation to their original dimensionality, ensuring compatibility with downstream fusion and perception tasks.

\subsection{Semantic Analog-to-Digital Converter}

\begin{figure*}[t!]
  
  \centering
  \includegraphics[width=0.92\linewidth]{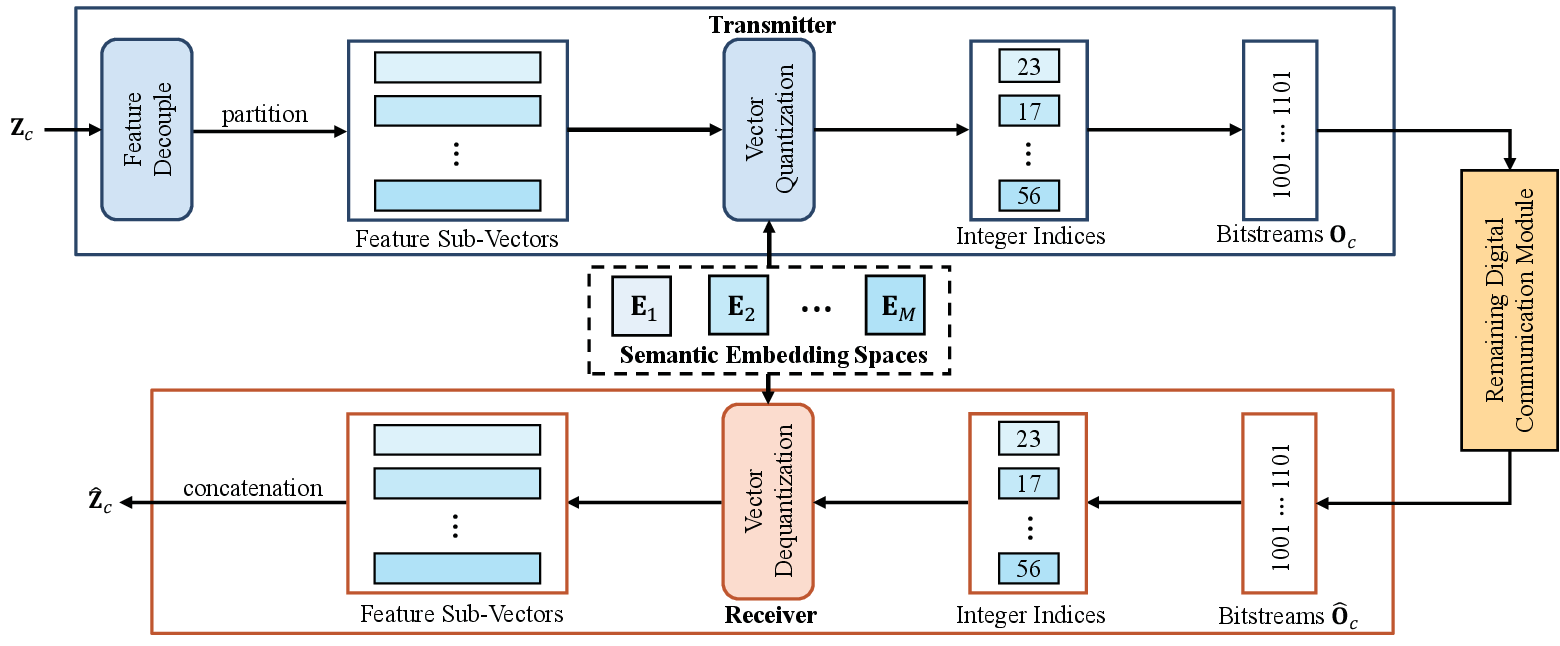}
  \caption{Network structure of our proposed semantic analog-to-digital converter.}  
  \label{fig:codebook}
\end{figure*}

In this subsection, we present the semantic analog-to-digital converter, denoted as $\mathcal{F}_{\omega}(\cdot)$ in (8), which enables bidirectional transformation between continuous semantic features and discrete bitstreams compatible with digital communication systems. 

The architecture of the semantic analog-to-digital converter is illustrated in Fig.~\ref{fig:codebook}.  
The converter contains two primary trainable components: a feature decoupling layer, denoted as $\mathbf{W}$, and a set of $M$ distinct semantic embedding spaces, collectively denoted as $\mathcal{E} = \{\mathbf{E}_1, \dots, \mathbf{E}_M\}$.
Each embedding space $\mathbf{E}_m = \{\mathbf{e}_{m,1}, \dots, \mathbf{e}_{m,N}\}$ contains $N$ semantic embedding vectors, with $\mathbf{e}_{m,n} \in \mathbb{R}^{L}$. To ensure coherent semantic representation, these embedding spaces are maintained consistently between the CAV and the ego vehicle.
 
At the transmitter,  we partition the $k$-th input semantic feature vector, denoted as $\mathbf{z}_k \in \mathbb{R}^{C' }$, into multiple sub-vectors to achieve a finer-grained representation of the semantic features. However, directly partitioning this vector is not feasible, as its components are inherently correlated. To address this, we first pass the vector through a feature decoupling layer to mitigate these correlations. The resulting decorrelated semantic feature vector is then partitioned into $M$ sub-vectors, forming the set $\{\mathbf{z}_k^{(1)}, \dots, \mathbf{z}_k^{(M)}\}$, where the length of each sub-vector, $L=C'/M$, matches that of the embedding vectors. Each sub-vector $\mathbf{z}_k^{(m)}$ is subsequently quantized by mapping it to the nearest embedding vector in its corresponding embedding space $\mathbf{E}_m$. This process yields a set of $M$ integer indices, which are concatenated and encoded into a binary bitstream for transmission.
 
At the receiver, the inverse operation is performed. The received bitstream is first decoded to recover the $M$ integer indices. Each recovered index serves as a pointer to retrieve the corresponding embedding vector from the locally synchronized semantic embedding space. Finally, these $M$ retrieved vectors are concatenated in their original order to reconstruct the complete semantic feature vector.

Specifically, to achieve a finer-grained representation, a strategy is to partition a feature vector into multiple sub-vectors. This approach can exponentially expand the model's expressive capacity; by selecting one of $N$ embedding vectors for each of the $M$ sub-vectors, a total of $N^M$ unique composite representations can be formed. As a result, more expressive and detailed semantic representations can be constructed by partitioning the feature vector. A straightforward implementation of this strategy is to directly partition the input semantic feature vector, $\mathbf{z}_k$.
However, the effectiveness of this direct partitioning scheme relies on the crucial assumption that feature dimensions can be organized into semantically independent sub-vectors.
This assumption is fundamentally inconsistent with the nature of features learned by the semantic compression encoder, which are highly entangled, meaning that semantic concepts are coupled and distributed across the entire feature vector. Consequently, naive partitioning of such features creates sub-vectors that lack complete or meaningful semantic content, leading to unstable training and significantly degraded model performance.

To resolve this issue, we introduce a feature decoupling layer that operates prior to the partitioning step. This layer is designed to transform the entangled features into a representation in which semantic concepts are disentangled, making them suitable for partitioning. Specifically, the layer learns a linear transformation matrix $\mathbf{W} \in \mathbb{R}^{C' \times C'}$ to map the input semantic feature $\mathbf{z}_k$ to a decoupled representation $\mathbf{z}'_k$, expressed as
\begin{equation}
    \mathbf{z}'_k = \mathbf{W} \mathbf{z}_k.
\end{equation}

To ensure the stability and effectiveness of this transformation, the matrix $\mathbf{W}$ is implemented without a bias term and is initialized as an identity matrix, i.e., $\mathbf{W}_{\text{init}} = \mathbf{I}$. This initialization allows the layer to begin as an identity mapping and learns a pure basis transformation, which gradually disentangles the feature space during training. 
 
Following the feature decoupling, the vector $\mathbf{z}'_k$ is partitioned into $M$ sub-vectors. Each sub-vector, $\mathbf{z}_k^{(m)}$, is then independently quantized by its corresponding embedding space, $\mathbf{E}_m$. 
\textcolor{black}{Specifically, for each sub-vector, we identify the index $I_{k,m}$ of the most semantically similar embedding vector from the corresponding embedding space. This is achieved by maximizing their cosine similarity, as expressed by
\begin{equation}
   \label{eq:quantization_index}
   I_{k,m} = \operatorname*{arg\,max}\limits_{n \in \{1, \dots, N\}} \text{cosine}(\mathbf{z}_k^{(m)}, \mathbf{e}_{m,n}),
\end{equation}
where the cosine similarity function, $\text{cosine}(\cdot, \cdot)$, is defined as the normalized inner product, formulated as
\begin{equation}
   \label{eq:cosine_similarity_def}
   \text{cosine}(\mathbf{a}, \mathbf{b}) = \frac{ \mathbf{a}^{\top} \mathbf{b} }{ \| \mathbf{a} \|_2 \| \mathbf{b} \|_2 },
\end{equation}
where $\| \cdot \|_2$ denotes the Euclidean norm.}
The resulting set of integer indices $\{I_{k,1}, I_{k,2}, \dots, I_{k, M}\}$ for the $k$-th input semantic feature vector $\mathbf{z}_k$ is subsequently encoded into a binary bitstream $\mathbf{O}_c$, and fed into the digital encoder.

The semantic embedding space is dynamically optimized throughout the training process. As part of the network’s end-to-end training, the active embedding vectors—those selected during the quantization step—are updated via gradient descent~\cite{vqvae}. This optimization adjusts each selected embedding vector to move closer to the feature sub-vector it represents, thereby minimizing the distance between them. Consequently, the embedding vectors are refined in the feature space, enhancing their ability to capture the underlying semantic structure of the data.
However, a notable limitation of this gradient-based approach is the potential for low embedding vector utilization. In practice, only a small subset of vectors may be consistently selected and updated, leaving a significant portion of the embedding space dormant. This phenomenon severely constrains the model's expressive capacity, limiting its ability to represent diverse and complex semantic structures present in the input data.
To overcome this limitation, we adopt a clustering-based vector quantization method. In addition to updating the embedding vectors via gradient descent, this method continuously monitors embedding vector utilization during training and strategically optimizes underutilized vectors to enhance overall efficiency~\cite{cvq}. The detailed reinitialization process is described as follows.
 
First,  the utilization rate of each semantic embedding vector is tracked across batches using an exponential moving average. To simplify the presentation, we denote the $n$-th vector in any semantic embedding space as $\mathbf{e}_n$. The utilization rate of the $n$-th embedding vector, $\mathbf{e}_n$, at batch $t$ is updated as
\begin{equation}
R_n^{(t)} = \rho R_n^{(t-1)} + (1 - \rho) \frac{r_n^{(t)}}{B \times K},
\end{equation}
where $R_n^{(t)}$ and $R_n^{(t-1)}$ are the utilization rates at the current batch $t$ and the previous batch $t-1$, respectively. The term $r_n^{(t)}$ denotes the number of times $\mathbf{e}_n$ is selected in the current batch. This count is normalized by the product of the batch size, $B$, and the number of semantic features per sample, $K$. The hyperparameter $\rho \in [0, 1)$ controls the decay rate of the moving average.

Next, an anchor-based update mechanism optimizes the embedding vectors. At each update step, a set of $V$ anchor vectors, $\tilde{\mathbf{Z}}^{(t)} = \{\tilde{\mathbf{z}}_1^{(t)}, \tilde{\mathbf{z}}_2^{(t)}, \ldots, \tilde{\mathbf{z}}_V^{(t)}\}$, is randomly sampled from semantic features $\mathbf{Z}_{c}$. For each embedding vector $\mathbf{e}_n$ that requires reinitialization, a corresponding anchor vector, $\tilde{\mathbf{z}}_v^{(t)}$, is probabilistically selected from $\tilde{\mathbf{Z}}^{(t)}$. The probability of selecting anchor $\tilde{\mathbf{z}}_v^{(t)}$ to update $\mathbf{e}_n$ is computed using a softmax function over the cosine similarity, $D_{n,v}^{(t)} =  \text{cosine}(\tilde{\mathbf{z}}_v^{(t)}, \mathbf{e}_n)$, as follows
\begin{equation}
p_{n,v}^{(t)} = \frac{\exp \left( -D_{n,v}^{(t)} \right)}{\sum_{v=1}^{V} \exp \left( -D_{n,v}^{(t)} \right)}.
\label{eq:sampling_prob}
\end{equation}

Finally, the embedding vector $\mathbf{e}_n$ is updated by optimizing it as a weighted average of its previous state and the selected anchor vector, as follows
\begin{equation}
\mathbf{e}_n^{(t)} = (1 - w_n^{(t)}) \mathbf{e}_n^{(t-1)} + w_n^{(t)} \tilde{\mathbf{z}}_v^{(t)}.
\label{eq:final_update}
\end{equation}
where $w_n^{(t)}$ is the update weight, determined by the vector's utilization rate, expressed as
\begin{equation}
w_n^{(t)} = \exp \left( -\frac{R_n^{(t)} \times N \times 10}{1 - \rho} - \epsilon \right),
\label{eq:update_weight}
\end{equation}
where $\epsilon$ is a small constant for numerical stability. This formulation ensures that embedding vectors with low utilization rates receive update weights approaching $1$, enabling significant updates, while frequently used vectors receive weights close to $0$, thereby preserving their learned positions.

\subsection{Training Strategy}

We train the proposed model using a four-stage strategy for stable convergence and desirable performance. The process begins with the independent training of the semantic compression codec on the perception task. This is followed by the separate training of the analog-to-digital converter, which learns to process the features generated by the frozen, pre-trained codec.  With these modules pre-trained, the entire network undergoes joint end-to-end training. The final stage is the task-oriented fine-tuning to optimize performance on the downstream task.

The first stage is dedicated to optimizing the semantic compression codec, $\mathcal{F}_{\theta}(\cdot)$ and $\mathcal{F}_{\phi}(\cdot)$, while all other network components are frozen, with their weights initialized from a publicly available pre-trained implementation. During this stage, the model is trained in an end-to-end manner, with the digital communication module bypassed. The input point cloud data is processed through the entire perception pipeline: from feature extraction and selection, through the semantic compression encoder and decoder, and finally to the fusion and detection network.
The loss function at this stage is designed for the 3D object detection task, aiming to enhance detection accuracy. Following the PointPillars~\cite{pointpillars}, the loss function combines classification and regression terms, and is formulated as
\begin{equation}
\mathcal{L}_{\text{stage1}} = \mathcal{L}_{\text{cls}}(\mathbf{\hat{Y}}, \mathbf{Y}) + \eta \cdot \mathcal{L}_{\text{reg}}(\mathbf{\hat{Y}}, \mathbf{Y}),
\label{eq:loss_stage1}
\end{equation} 
where $\mathcal{L}_{\text{cls}}$ denotes focal loss~\cite{focalloss} for classification, $\mathcal{L}_{\text{reg}}$ represents smooth $\ell_1$ loss for regression, $\eta$ is a weighting factor, and $\mathbf{Y}$ denotes the ground-truth detection results.

The second stage focuses exclusively on optimizing the semantic analog-to-digital converter, $\mathcal{F}_{\omega}(\cdot)$. In this stage, the converter is inserted between the pre-trained semantic compression encoder and decoder derived in the first stage, while all other network components remain frozen. The objective of this stage is to train the converter to perform high-fidelity quantization and reconstruction of the semantic features, $\mathbf{Z}_c$. To this end, the optimization is guided by a reconstruction loss, defined as the mean squared error (MSE) between the original semantic features, $\mathbf{Z}_c$, and the reconstructed features, $\mathbf{\hat{Z}}_c$, expressed as
\begin{equation}
\mathcal{L}_{\text{stage2}} = \mathcal{L}_{\text{mse}}(\mathbf{\hat{Z}}_c, \mathbf{Z}_c),
\label{eq:loss_stage2}
\end{equation}
where $\mathcal{L}_{\text{mse}}(\cdot)$ denotes the MSE loss. To enable gradient flow through the non-differentiable quantization operations in the semantic analog-to-digital converter, the straight-through estimator (STE) is employed to approximate gradients during backpropagation~\cite{ste}.

In the third stage, the entire network is jointly trained in an end-to-end manner, including $\mathcal{F}_{\alpha}(\cdot)$, $\mathcal{F}_{\beta}(\cdot)$, $\mathcal{F}_{\theta}(\cdot)$, $\mathcal{F}_{\phi}(\cdot)$, $\mathcal{F}_{\kappa}(\cdot)$, and $\mathcal{F}_{\omega}(\cdot)$. The training loss combines the task-oriented detection loss from the first stage and the reconstruction loss from the second stage, defined as
\begin{equation}
\mathcal{L}_{\text{stage3}} = \mathcal{L}_{\text{cls}}(\mathbf{\hat{Y}}, \mathbf{Y}) + \eta \cdot \mathcal{L}_{\text{reg}}(\mathbf{\hat{Y}}, \mathbf{Y}) + \lambda \cdot \mathcal{L}_{\text{mse}}(\mathbf{\hat{Z}}_c, \mathbf{Z}_c),
\label{eq:stage3_loss}
\end{equation}
where $\lambda$ is a hyperparameter that balances the detection and reconstruction losses.

The final stage consists of end-to-end fine-tuning to align the decoding and perception pipeline with the object detection task. In this stage, the semantic compression decoder $\mathcal{F}_{\phi}(\cdot)$, the semantic analog-to-digital converter $\mathcal{F}_{\omega}(\cdot)$ (excluding its feature decoupling layer $\mathbf{W}$), and the detection network $\mathcal{F}_{\kappa}(\cdot)$ are jointly optimized, while other network components remain frozen. The optimization is driven exclusively by the task-oriented detection loss. This process refines the representations learned by the converter and decoder, ensuring they are maximally informative for the final perception task. The objective function is therefore identical to that used in the first stage, and is defined as
\begin{equation}
\mathcal{L}_{\text{stage4}} = \mathcal{L}_{\text{cls}}(\mathbf{\hat{Y}}, \mathbf{Y}) + \eta \cdot \mathcal{L}_{\text{reg}}(\mathbf{\hat{Y}}, \mathbf{Y}).
\label{eq:stage4_loss}
\end{equation}

\subsection{UAN Design and Loss Function}

This subsection presents the design of the UAN and its corresponding loss function, developed to mitigate performance collapse in digital communication systems operating at low SNR.
 
In digital communication systems, standard hard-decision channel decoding is inherently fragile for semantic representation. A single bit error in the transmitted binary representation of an embedding vector can cause the system to select another semantically distant vector from the embedding space, severely distorting the reconstructed meaning. This leads to the cliff effect, a catastrophic collapse in performance once channel noise exceeds a critical threshold.

To address this challenge, we introduce two key strategies. 
First, to mitigate the impact of bit errors, we employ a Gray code design for the semantic embedding space. This involves projecting high-dimensional embedding vectors onto a 2D plane using an averaging operation and applying a Gray code to their spatial arrangement. This structure ensures that a single-bit error in a transmitted index yields a spatially proximate and thus semantically similar embedding vector, thereby minimizing distortion.
Second, to address the unreliability of features decoded under low SNR conditions, we leverage the principle that erroneous features degrade overall perception performance more severely than missing ones. Specifically, we propose the UAN to assess the reliability of each reconstructed semantic feature $\mathbf{\hat{z}}_k$ and output a corresponding confidence score $p_k$. Using these confidence scores, the system can selectively discard features deemed unreliable, thereby preventing error propagation and ensuring graceful performance degradation under low-SNR conditions.

\begin{figure}[t!]
  
  \centering
  \includegraphics[width=3.5in]{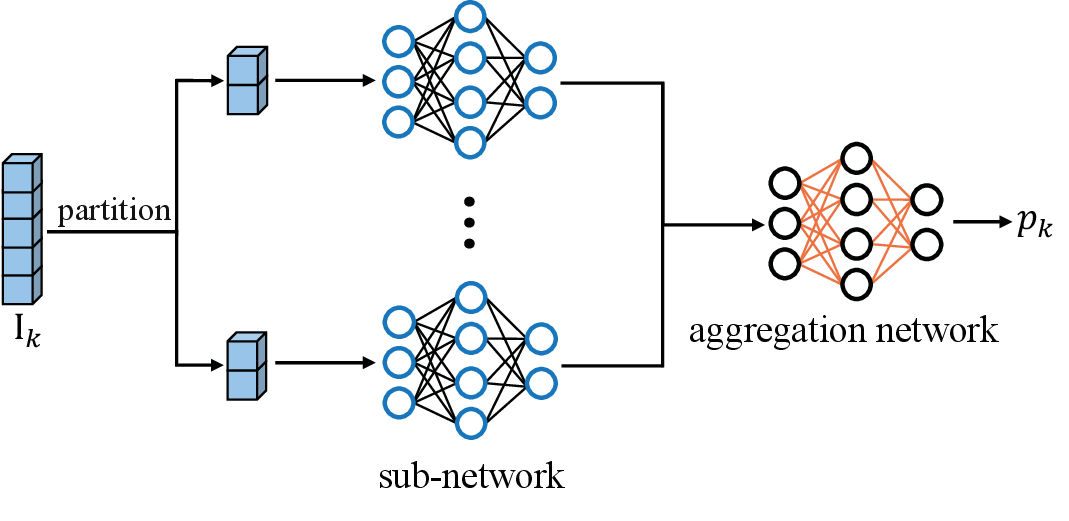}
  \caption{Network structure of our proposed UAN.}  
  \label{fig:UAN}
\end{figure}

As illustrated in Fig.~\ref{fig:UAN}, the UAN architecture is organized hierarchically to align with the structure of the multi-semantic embedding space. The network input is the LLR vector, denoted as $\mathbf{l}_k$, corresponding to the $k$-th reconstructed semantic feature, denoted as $\mathbf{\hat{z}}_k$. The LLR vector is first partitioned into $M$ groups and independently processed by an identical sub-network, composed of several linear layers, to produce an intermediate reliability score. These $M$ intermediate scores are then fed into a final aggregation network, which generates the confidence score $p_k$ for the $k$-th reconstructed semantic feature. 

A straightforward approach to training the UAN is to frame the task as a binary classification problem, predicting the correctness of a decoded feature using the binary cross-entropy (BCE) loss \cite{bce}. However, BCE assigns equal penalties to all decoding errors, which reduces efficiency in our setting since different errors have markedly different semantic impacts under the adopted Gray coding scheme.
For example, a single-bit error causing a shift to an adjacent index results in far less semantic damage than a multi-bit error.
To this end, we propose a semantic-weighted binary cross-entropy (SW-BCE) loss. 
This loss function modulates the standard BCE penalty for each incorrectly decoded feature based on its associated semantic error. This trains the UAN to be more sensitive to the most semantically destructive errors, rather than treating all failures equally.

Specifically, a binary ground-truth label, $y_k \in \{0, 1\}$, is used to indicate whether the embedding vector of the $k$-th reconstructed feature, $\mathbf{\hat{z}}_k$, has been correctly decoded after transmission. This label is defined as
\begin{equation}
y_k =
\begin{cases}
1, & \text{if } \mathbf{\hat{z}}_k = \mathbf{c}_k, \\
0, & \text{otherwise},
\end{cases}
\label{eq:ground_truth}
\end{equation}
where $\mathbf{c}_k$ is the concatenated representation of the $M$ embedding vectors corresponding to the $k$-th semantic feature on the CAV side.
Furthermore, to quantify the magnitude of the semantic error, we introduce a semantic damage metric, denoted as $d_k$. This metric is defined as 
\begin{equation}
   d_k = 1 -  \text{cosine}(\mathbf{c}_k, \mathbf{\hat{z}}_{k}).
   \label{eq:semantic_damage}
\end{equation}

\begin{algorithm}[t]
\caption{Training the UAN with the SW-BCE Loss}
\label{alg1}
\begin{algorithmic}[1]
\STATE \textbf{Input:} Training dataset $\mathcal{D}$; hyperparameter $\alpha$.
\STATE \textbf{Initialize:} Load and freeze the pre-trained network parameters $\omega$; initialize UAN parameters $\varphi$ randomly.
\STATE \textbf{while} not converged \textbf{do}
\STATE \quad Sample a mini-batch of $\{\mathbf{z}_k\}_{k=1}^K$ from $\mathcal{D}$.
\STATE \quad \textbf{for} $k=1, \dots, K$ \textbf{do}
\STATE \quad\quad Compute LLR vector $\mathbf{l}_k$ using (8)-(11).
\STATE \quad\quad Predict confidence score $p_k$ using (13).
\STATE \quad\quad Determine the ground-truth label $y_k$ using (\ref{eq:ground_truth}).
\STATE \quad\quad Compute the semantic damage $d_k$ using (\ref{eq:semantic_damage}).
\STATE \quad\quad Calculate the sample-wise loss $\mathcal{L}_{\text{SW-BCE}}$ via (\ref{eq:sw_bce_loss}).
\STATE \quad \textbf{end for}
\STATE \quad Compute the total mini-batch loss $\mathcal{L}_{\text{SW-BCE}}$ using (\ref{eq:total_loss}).
\STATE \quad Update UAN parameters $\varphi$ to minimize $\mathcal{L}_{\text{SW-BCE}}$ via \\ \quad  gradient descent.
\STATE \textbf{end while}
\STATE \textbf{Return:} Trained UAN parameters $\varphi$.
\end{algorithmic}
\end{algorithm}

For correctly decoded features with $y_k = 1$, the semantic damage is zero, i.e., $d_k = 0$. Leveraging these components, the SW-BCE loss for the $k$-th feature, denoted as $\mathcal{L}_{\text{SW-BCE},k}$, is formulated as 
\begin{equation}
\mathcal{L}_{\text{SW-BCE},k} = -[y_k \log(p_k) + (1-y_k)(1 + \alpha d_k) \log(1-p_k)],
\label{eq:sw_bce_loss}
\end{equation}
where $\alpha \geq 0$ is a hyperparameter that controls the scaling of the penalty based on semantic damage. In the case of a correctly decoded feature, the loss reduces to the standard BCE loss. Conversely, when a decoding error occurs, the loss is amplified by a factor of $1 + \alpha d_k$. This mechanism compels the network to develop a heightened sensitivity to the most semantically significant errors, thereby enhancing its ability to identify and suppress the most harmful decoding artifacts.

Finally, the total loss for training the UAN is the average of the SW-BCE loss across all $K$ semantic features, given by 
\begin{equation}
\mathcal{L}_{\text{SW-BCE}} = \frac{1}{K} \sum_{k=1}^{K} \mathcal {L}_{\text{SW-BCE},k}.
\label{eq:total_loss}
\end{equation}

The training of the UAN is solely implemented after training the semantic compression codec and the semantic analog-to-digital converter. The detailed training process is outlined in Algorithm~\ref{alg1}.

\section{Simulation Results}	
This section provides a comprehensive performance evaluation of our proposed CoDS framework.

\subsection{Experimental Setup}
 
\textbf{Dataset.}
To evaluate the performance of the proposed collaborative perception framework, we employed OPV2V, a publicly available large-scale dataset specifically designed for V2V collaborative perception tasks~\cite{opencood}. Generated through the OpenCDA platform that integrates CARLA and SUMO simulators~\cite{carla}, the OPV2V dataset comprises 12,000 synchronized frames containing both LiDAR point clouds and RGB images, encompassing 230,000 meticulously annotated 3D bounding boxes. Following the standard experimental protocol, we adopted the official data partition: 6,764 frames for training, 1,981 frames for validation, and 2,170 frames for testing. For 3D object detection evaluation, the perception region is defined as $x \in [-140, 140]$~meters and $y \in [-40, 40]$~meters, centered on the ego vehicle. To simulate realistic communication constraints, the maximum communication range is set to 70 meters between the ego vehicle and collaborating agents.

\textbf{Implementation Details.}
The proposed model is implemented in PyTorch and utilizes the Adam optimizer with an initial learning rate of $1 \times 10^{-4}$ and an exponential weight decay factor of 0.8, training the network until convergence.

The architecture of our digital semantic communication system is configured as follows. The semantic compression codec achieves a $\gamma_c = 16$ using $N_{b} = 4$ stacked down/up-sampling blocks. The semantic analog-to-digital converter component utilizes $M = 4$ semantic embedding spaces, each containing $N = 64$ embedding vectors of length $L = 16$. Besides, the system employs an adaptive modulation and coding (AMC) scheme to dynamically select the optimal modulation and coding scheme (MCS) according to the channel conditions. The set of available MCSs is designed with reference to the IEEE 802.11bd standard~\cite{standard}, implementing LDPC codes with a block length of $1,296$. Finally, the UAN is trained exclusively over a Rayleigh fading channel.

\textbf{Benchmarks.} We evaluate the proposed CoDS scheme against three benchmarks, all of which adopt PointPillars as the underlying backbone. The baselines are as follows
\begin{itemize}
    \item \textbf{CNN-based ASC:} This baseline is adapted from the ASC for collaborative perception scheme in~\cite{franklin}. It utilizes a convolutional neural network (CNN)-based analog semantic codec network to transmit features over noisy channels.
    \item \textbf{Transformer-based ASC:} This baseline is adapted from the approach in~\cite{scomcp} and utilizes a Transformer-based analog semantic codec network to perform the same task.
    \item \textbf{Traditional Digital Communication:} The non-semantic baseline is implemented using a standard digital pipeline that supports multiple MCSs. The channel coding component utilizes LDPC codes with a block length of $1,296$.
\end{itemize}

\begin{figure*}[t]
  \centering
  \begin{minipage}[b]{0.49\textwidth}
      \centering
      \includegraphics[width=\linewidth]{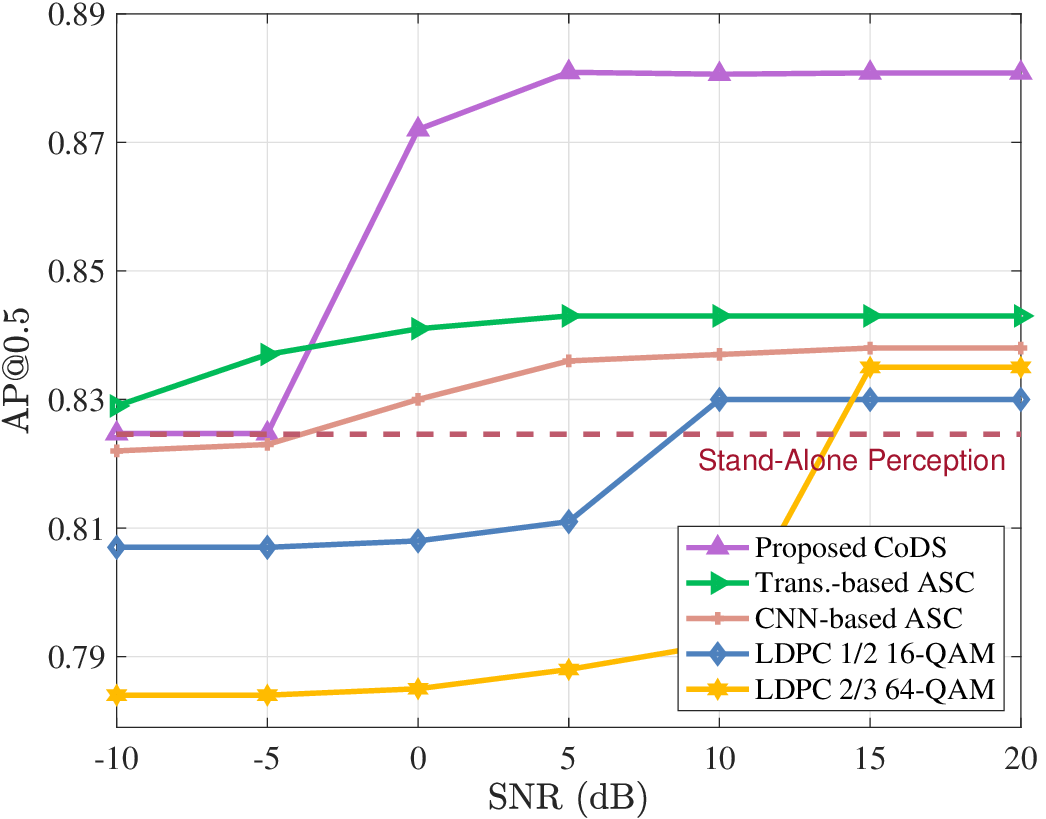}
      \centerline{(a) Perception performance in terms of $\text{AP@}0.5$.}
  \end{minipage}
  \hspace{0.005\textwidth}
  \begin{minipage}[b]{0.49\textwidth}
      \centering
      \includegraphics[width=\linewidth]{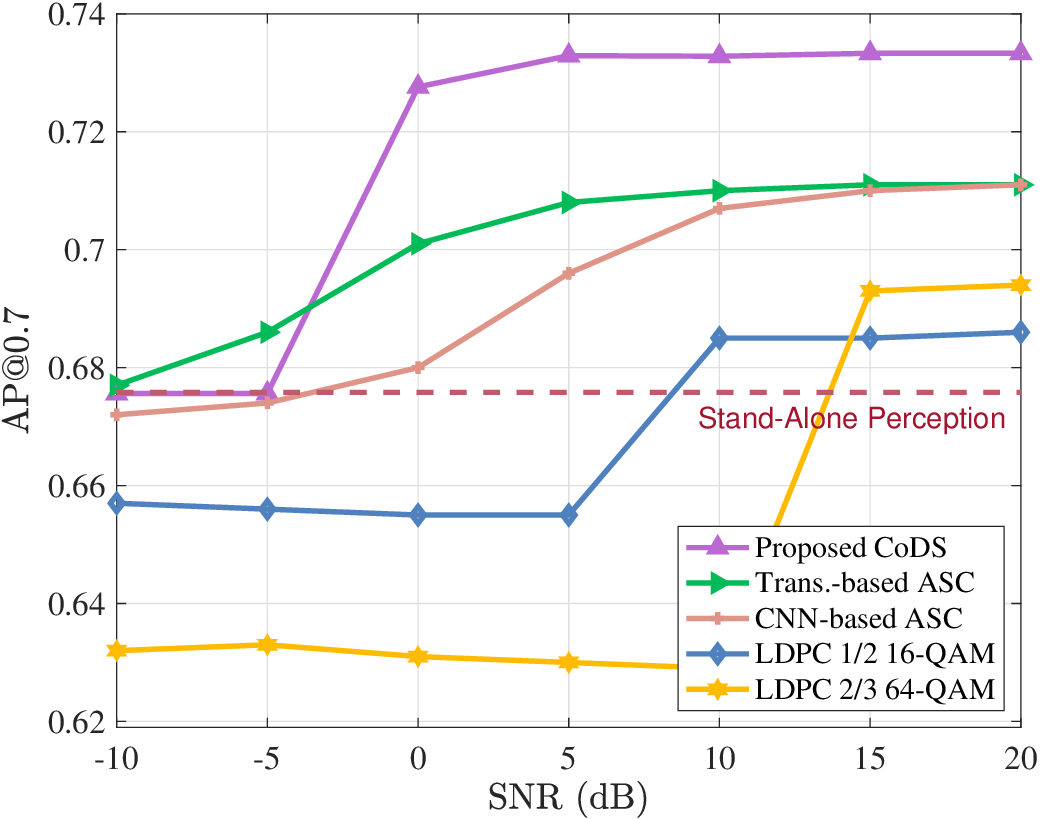}
      \centerline{(b) Perception performance in terms of $\text{AP@}0.7$.}
  \end{minipage}
  \caption{Performance of the proposed method compared with other schemes at different SNRs over an AWGN channel, where “Trans.” denotes Transformer.}
  \label{fig:awgn}
\end{figure*}

  \begin{figure*}[h]
  \centering
  \begin{minipage}[b]{0.49\textwidth}
      \centering
      \includegraphics[width=\linewidth]{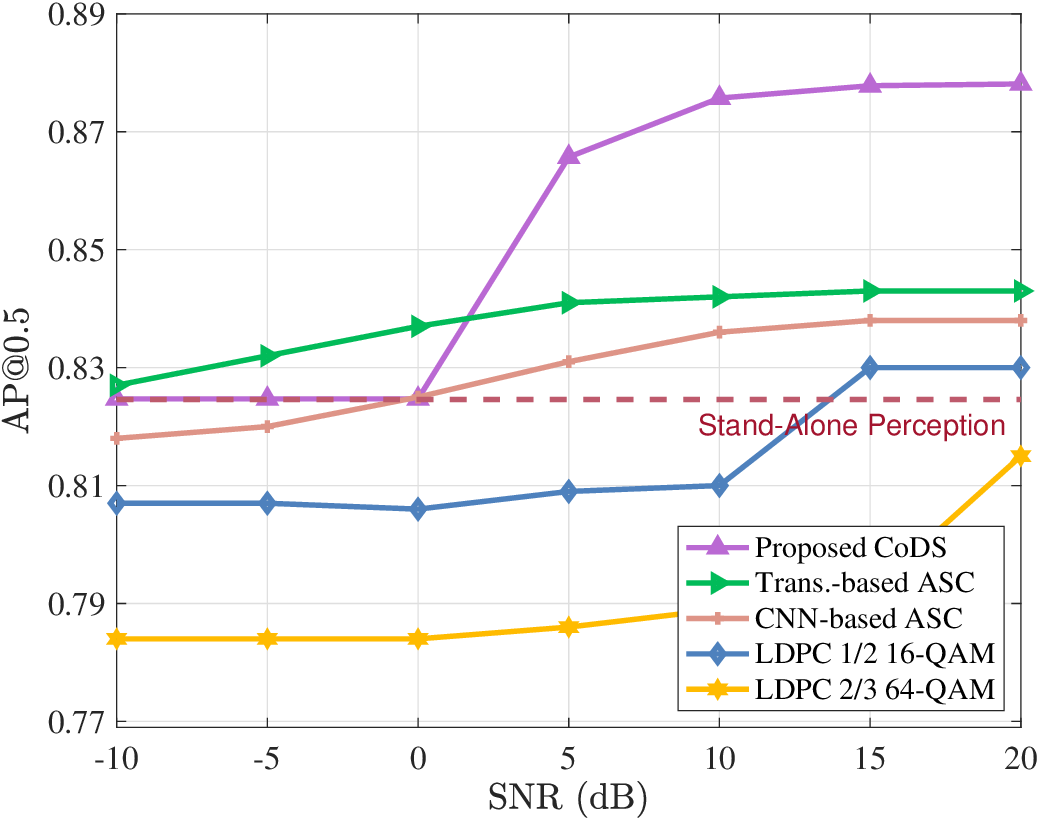}
      \centerline{(a) Perception performance in terms of $\text{AP@}0.5$.}
  \end{minipage}
  \hspace{0.005\textwidth}
  \begin{minipage}[b]{0.49\textwidth}
      \centering
      \includegraphics[width=\linewidth]{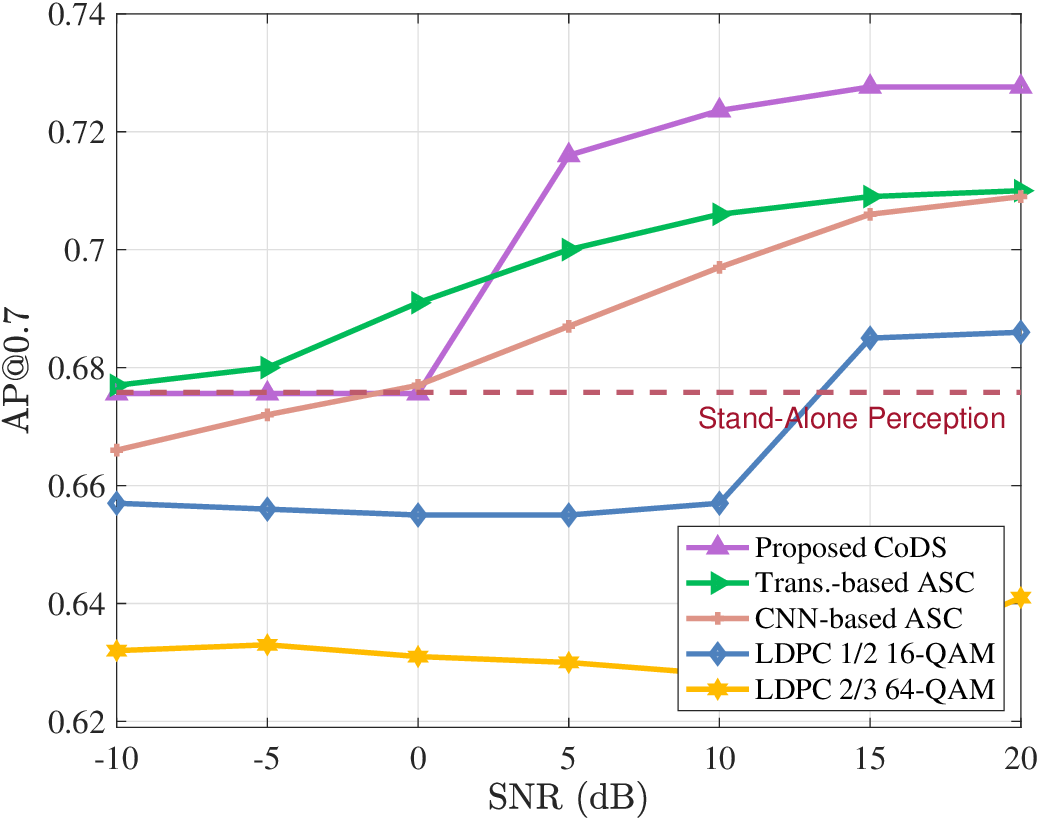}
      \centerline{(b) Perception performance in terms of $\text{AP@}0.7$.}
  \end{minipage}
  \caption{Performance of our proposed method compared with other schemes at different SNRs over a Rayleigh fading channel, where “Trans.” denotes Transformer.}
  \label{fig:rayleigh}
\end{figure*}

\subsection{Evaluation Metrics}

The performance is evaluated using average precision (AP), a standard metric for 3D object detection \cite{ap1}. This metric quantifies detection accuracy by calculating the area under the precision-recall curve, with higher values indicating better performance. 
We report AP at intersection-over-union (IoU) thresholds of 0.5 and 0.7, denoted as $\text{AP@}0.5$ and $\text{AP@}0.7$, respectively. Since higher IoU thresholds impose stricter detection criteria by requiring more precise object localization, performance at $\text{AP@}0.7$ is typically lower than that at $\text{AP@}0.5$.

The perception performance is highly dependent on the SNR of the physical noisy channel, which determines the reconstructed quality of the transmitted features. The SNR is defined as the ratio of the average power of the transmitted symbols, denoted as $P_s$, to the average power of the channel noise. Since the average power of each complex symbol is normalized to unity, we have $P_s = \mathbb{E}[|\mathbf{S}_c|^2] = D$. Accordingly, the SNR in decibels (dB) is expressed as
\begin{equation}
\label{eq:snr_definition}
\text{SNR (dB)} = 10 \log_{10} \left( \frac{P_s}{\sigma^2} \right),
\end{equation}

Different transmission approaches operating at a specific $\gamma_s$ may incur varying communication overhead when transmitting identical feature maps. 
To establish a fair evaluation framework, we normalize the communication cost by constraining all competing methods to utilize an identical number of channel uses, which represents the total number of complex symbols transmitted over the wireless channel.
For the proposed digital semantic communication system, the number of channel uses is given by
\begin{equation}
\label{eq:channel_uses_digital}
\text{Channel Uses} = \frac{Q_d \times q}{C \times \gamma_s \times R_c \times \log_2 O_r},
\end{equation}
where $Q_d = H \times W \times C$ denotes the data size of the feature map $\mathbf{M}_c$, $R_c$ is the LDPC code rate, and $ O_r$ is the modulation order (e.g., $ O_r=16$ for 16-QAM). 
Similarly, for the traditional digital communication approach, the number of channel uses is given by
\begin{equation}
\label{eq:channel_uses_traditional}
\text{Channel Uses} = \frac{Q_d  \times 8}{ \gamma_s \times R_c \times \log_2 O_r}.
\end{equation}

For the ASC scheme, the number of channel uses is calculated as
\begin{equation}
  \label{eq:channel_uses}
  \text{Channel Uses} =\frac{ Q_d }{ \gamma_s}.
\end{equation}

To ensure a fair comparison, the parameter $\gamma_s$ is adjusted until the calculated number of channel uses is identical across all methods.
  
\subsection{Performance Analysis}

We first present experimental results comparing our proposed CoDS framework with leading ASC methods and conventional digital communication baselines. Fig.~\ref{fig:awgn} illustrates the perception performance comparison between CoDS and other methods at different SNRs over an AWGN channel.  
A reference baseline are included: the performance of a stand-alone perception system without collaboration, shown as a red dotted line.
Compared to ASC schemes, at $\text{SNR} = -10$~dB, CoDS achieves a modest $0.28\%$ performance gain over the CNN-based ASC but is outperformed by the Transformer-based ASC.
This behavior stems from the inherent susceptibility of digital communication to the cliff effect, where performance degrades precipitously under severe channel conditions. 
The effect is evident when comparing modulation orders. Although higher-order constellations (e.g., 64-QAM) are spectrally efficient at high SNRs, their smaller minimum distance between constellation points makes them more noise-sensitive, leading to elevated bit error rates and frequent decoding failures. In collaborative perception, these failures propagate corrupted features to downstream tasks and can precipitate sudden accuracy collapse. Accordingly, at low SNRs, the dense 64-QAM constellation is prone to reconstruction failure, whereas a simpler scheme such as 16-QAM is more robust. Moreover, attempting to transmit more bits under such poor channel conditions is counterproductive: it increases the volume of corrupted features and further degrades perception performance. Although the proposed UAN mitigates these effects by filtering unreliable semantic features, some performance degradation at extremely low SNRs is unavoidable, causing CoDS to approach the stand-alone baseline.
In contrast, the ASC baselines, including both CNN-based and Transformer-based approaches, exhibit graceful performance degradation. Their performance declines smoothly as channel conditions worsen, demonstrating greater robustness to channel impairments compared with the brittle behavior of conventional digital systems. However, while they effectively mitigate the cliff effect, they still fall short of our proposed method at high SNRs. At $\text{SNR} = 20$~dB, CoDS surpasses the CNN-based and Transformer-based ASC schemes by notable margins of $4.25\%$ and $3.78\%$, respectively. This result underscores the strength of our digital semantic architecture in reliably transmitting rich and essential perceptual information under favorable channel conditions.
The superiority of CoDS arises from two complementary design principles. First, it adopts a robust digital semantic architecture. Semantic features are quantized into a bitstream and protected via channel coding, enabling fine-grained, bit-level protection of critical semantics and more reliable conveyance of rich, essential information within comparable or lower bandwidth. Second, CoDS employs dual adaptation to handle channel variation and residual errors. At the transmitter, AMC selects modulation and code rate matched to instantaneous SNR—using lower-order modulation at low SNR for reliability, and higher-order settings at high SNR for throughput. At the receiver, the proposed UAN evaluates the reliability of decoded semantic features and discards those deemed corrupted, limiting error propagation and mitigating the cliff effect. Ultimately, this holistic design achieves a superior trade-off between perception accuracy and communication cost, establishing CoDS as a more effective and robust solution for reliable collaborative perception in noisy wireless environments.
As shown in Fig.~\ref{fig:awgn}(b), we compare the performance of the proposed method with other schemes in terms of $\text{AP@}0.7$. Consistent with the findings at $\text{AP@}0.5$, our method significantly outperforms other schemes under high SNR conditions, while its performance in the low SNR regime converges toward that of the stand-alone bound. As expected, all schemes exhibit a marked performance drop when evaluated at this higher IoU threshold. The stricter criterion for $\text{AP@}0.7$ demands greater precision in object localization, making the perception performance more susceptible to information errors induced by channel noise.

\begin{figure*}[t]
  \centering
  \begin{minipage}[b]{0.49\textwidth}
      \centering
      \includegraphics[width=\linewidth]{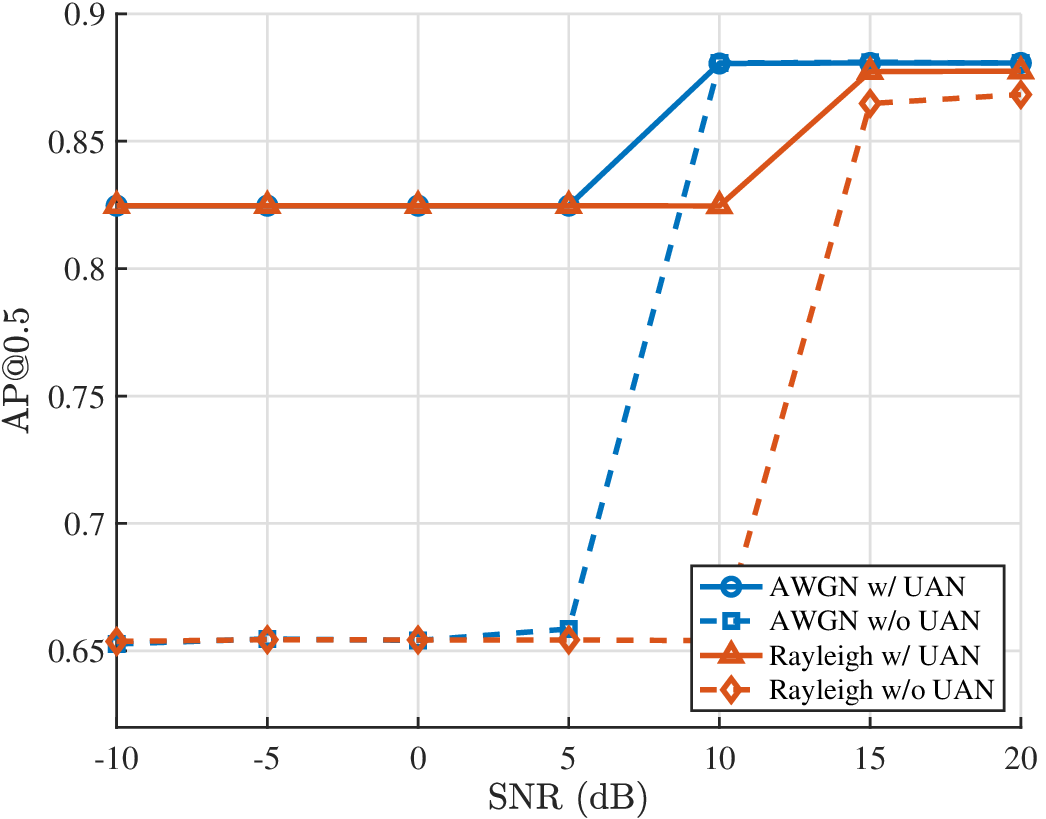}
      \centerline{(a) Perception performance in terms of $\text{AP@}0.5$.}
  \end{minipage}
  \hspace{0.005\textwidth}
  \begin{minipage}[b]{0.49\textwidth}
      \centering
      \includegraphics[width=\linewidth]{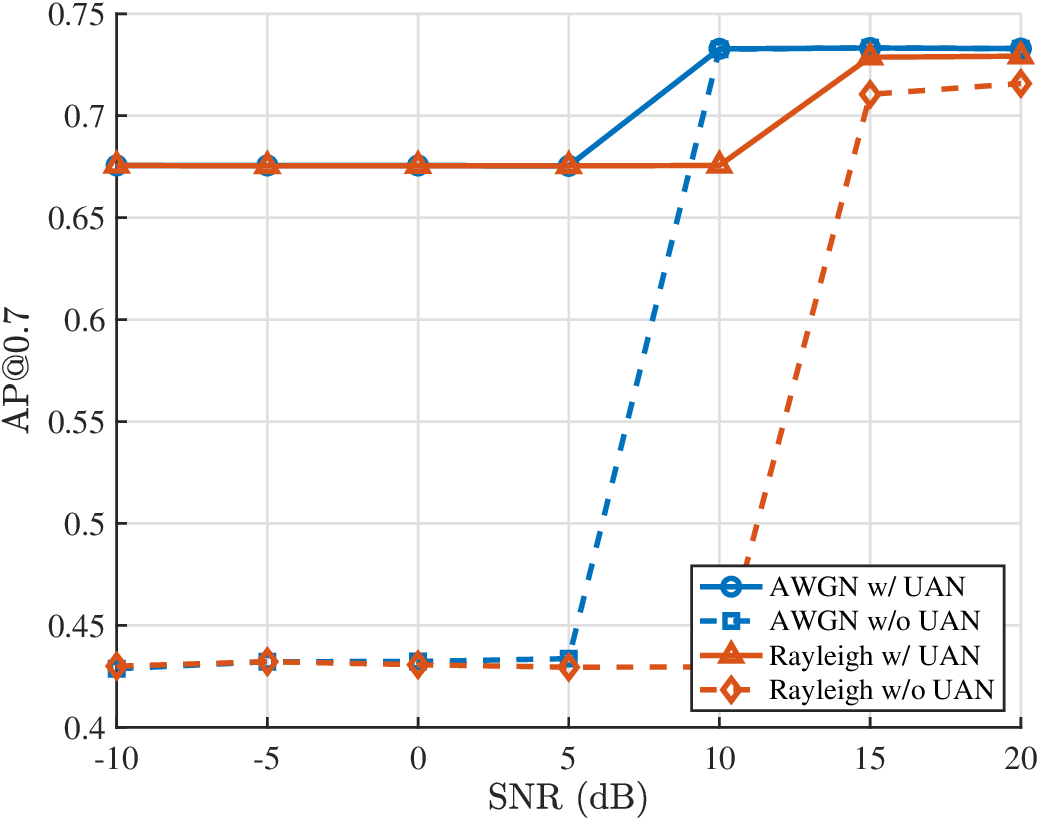}
      \centerline{(b) Perception performance in terms of $\text{AP@}0.7$.}
  \end{minipage}
  \caption{Performance of our proposed UAN at different SNRs over both AWGN and Rayleigh fading channels.}
  \label{fig:ablation_study_uncertainty}
\end{figure*}

\begin{table}[h!]
\centering
\begin{threeparttable}
\caption{Performance of proposed semantic compression codec at different $\gamma_c$ values.}
\label{tab2_transposed} 
\sisetup{detect-weight=true}
\setlength{\tabcolsep}{5.3pt}
\begin{tabular}{l *{4}{S[table-format=1.4]}}
\toprule
\textbf{} & { {No Compress}} & { {$\gamma_c$ = 4}} & { {$\gamma_c$ = 8}} & { {$\gamma_c$ = 16}} \\
\midrule
 {AP@0.5} & 0.8697 & 0.8700 & 0.8715 & 0.8702 \\
 
 {AP@0.7} & 0.7606 & 0.7598 & 0.7533 & 0.7534 \\
\bottomrule
\end{tabular}
\sisetup{detect-weight=false}
\end{threeparttable}
\end{table}

Fig.~\ref{fig:rayleigh} illustrates a comparison of perception performance between our proposed method and other schemes in terms of $\text{AP@}0.5$ and $\text{AP@}0.7$ at different SNRs over a Rayleigh fading channel. The multipath fading induces general performance degradation for all schemes compared to the AWGN case, and the relative performance trends across different SNRs remain consistent with those observed in Fig.~\ref{fig:awgn}. Crucially, although our model is trained solely over a Rayleigh fading channel, CoDS maintains competitive accuracy when evaluated on both Rayleigh and AWGN test scenarios without model retraining. This resilience to distinct channel conditions underscores the practical viability of our approach for deployment in dynamic real-world vehicular environments.

Table~\ref{tab2_transposed} presents the perception performance of the proposed semantic compression codec under different $\gamma_c$ values, with $\gamma_s = 3.3 \times 10^{2}$. The evaluation is performed under lossless transmission, excluding the digital communication module.
As shown in Table~\ref{tab2_transposed}, the proposed semantic compression codec achieves strong performance in terms of $\text{AP@}0.5$. Remarkably, at $\gamma_c$ values of $4$, $8$, and $16$, its performance is comparable to—or even exceeds—that of the uncompressed baseline. This improvement arises from the autoencoder-based architecture, which effectively refines and distills more informative semantic representations from the input features.
Under the stricter $\text{AP@}0.7$ metric, the network exhibits only marginal degradation. Therefore, we set $\gamma_c$ to $16$ in the experiment.

Fig.~\ref{fig:ablation_study_uncertainty} presents an ablation study to evaluate the proposed UAN at different SNRs over both AWGN and Rayleigh fading channels. In this experiment, for the digital codec, we employ a 1/2-rate LDPC code, and the resulting bits are modulated using 16-QAM.
The results demonstrate that the UAN effectively mitigates the characteristic cliff effect at low SNRs and provides performance improvements even at high SNRs.
Specifically, as illustrated in Fig.~\ref{fig:ablation_study_uncertainty}(a), the UAN demonstrates a substantial capacity to prevent abrupt performance degradation under challenging channel conditions in terms of $\text{AP@}0.5$. For instance, at an $\text{SNR} = -10$~dB, the UAN achieves a performance gain of $17.1\%$ compared to the baseline without this module over the Rayleigh fading channel. This improvement stems from the UAN's adaptive ability to identify and discard unreliable semantic features, ensuring graceful performance degradation rather than catastrophic failure. Even under favorable high SNR conditions, the UAN continues to provide performance improvements. For example, in the Rayleigh fading channel at $\text{SNR} = 20$~dB, it improves performance by $0.92\%$. 
The advantage is more pronounced under the stricter $\text{AP@}0.7$ metric. As shown in Fig.~\ref{fig:ablation_study_uncertainty}(b), the performance disparity between the UAN and the baseline without the UAN is particularly significant over the Rayleigh fading channel, where the UAN outperforms the baseline by $24.56\%$ at $\text{SNR} = -10$~dB and by $1.34\%$ at $\text{SNR} = 20$~dB in terms of $\text{AP@}0.7$. Collectively, these findings validate the role of the UAN as a critical component for building robust digital semantic communication systems. By preventing catastrophic performance collapse in low SNR regimes, the UAN ensures a high level of perceptual reliability, a crucial requirement for safety-critical applications such as autonomous driving.

\begin{figure}[t!]
  \centering
  \includegraphics[width=3.5in]{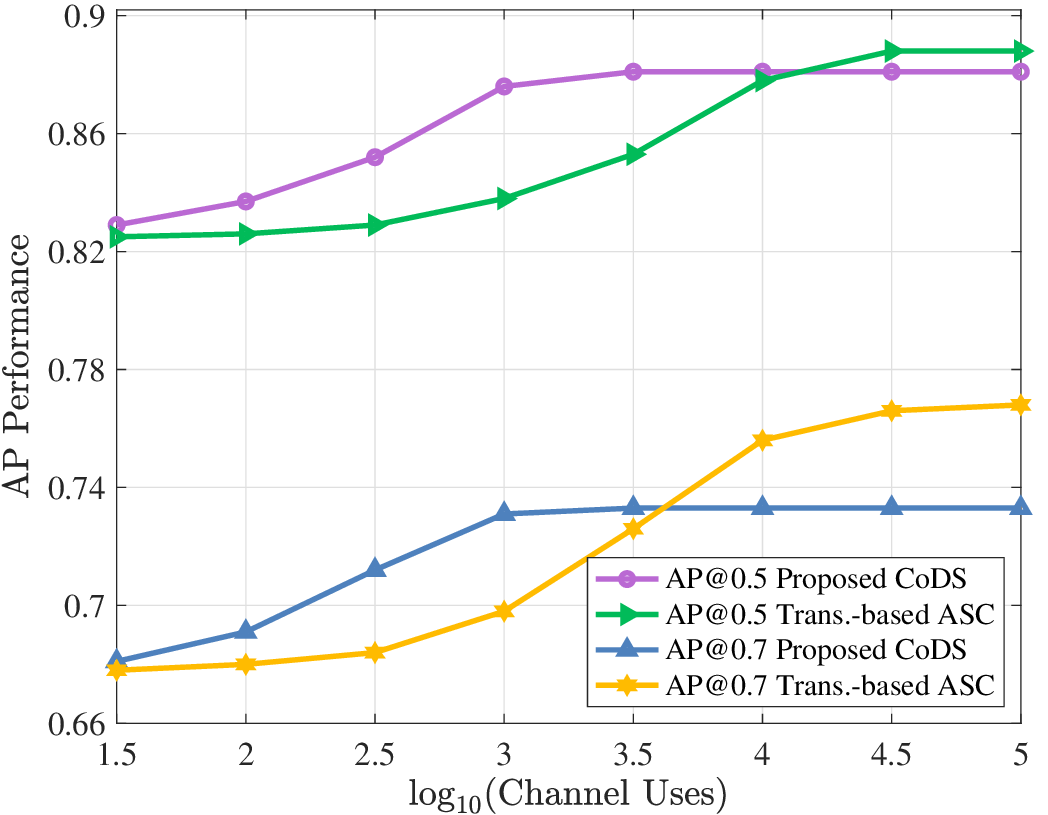}
  \caption{Performance of proposed CoDS compared with ASC scheme at different channel uses at $\text{SNR} = 5$~dB over an AWGN channel in terms of $\text{AP@}0.7$ and $\text{AP@}0.5$, where “Trans.” denotes Transformer.}
  \label{fig:channel_use}
\end{figure}

Fig.~\ref{fig:channel_use} compares the perception performance of the proposed CoDS framework and the Transformer-based ASC scheme under varying channel uses at $\text{SNR} = 5$~dB over an AWGN channel. As shown, the perception performance of both methods generally improves with increasing channel use.
In resource-constrained conditions (i.e., with few channel uses), CoDS outperforms the ASC baseline, demonstrating that the proposed digital semantic framework is more effective at transmitting essential information when communication resources are limited. However, as channel uses increase beyond a logarithmic value of approximately $3.5$, the performance of CoDS begins to saturate, while the ASC method continues to improve and eventually surpasses CoDS in terms of $\text{AP@}0.7$. This performance ceiling arises from the intrinsic representational limitations of the semantic embedding space in CoDS. In contrast, the analog nature of ASC enables it to exploit the additional bandwidth to convey progressively more perceptual details, resulting in higher performance when resources are abundant.
A similar trend is observed for the $\text{AP@}0.5$ metric, although the crossover occurs later—around a log value of $4.5$. This delay occurs because the core semantic information captured within the embedding space remains sufficient for the more lenient $\text{AP@}0.5$ detection threshold. The finer-grained details omitted by the embedding space have less impact on this metric, allowing CoDS to sustain its advantage longer.
Overall, these findings demonstrate that CoDS excels in practical, bandwidth-limited communication environments, where efficiently transmitting the most salient semantic information is of paramount importance. 

\section{Conclusion}

In this paper, we have proposed CoDS, a novel collaborative perception framework based on digital semantic communication. The framework enables efficient and robust semantic-level information sharing among CAVs while remaining compatible with existing digital communication infrastructures. Specifically, we first design a semantic compression codec to extract task-oriented features, effectively reducing communication overhead while preserving perception performance. Next, we introduce a semantic analog-to-digital converter to transform continuous semantic features into discrete bitstreams compatible with standard transmission pipelines. To further enhance reliability under adverse wireless conditions, we develop the UAN, which evaluates the reliability of received features and selectively filters them to mitigate the cliff effect at low SNRs. Experimental results show that CoDS outperforms both traditional digital communication and existing ASC schemes.

\bibliographystyle{IEEEtran}
\bibliography{IEEEabrv,reference}

\vfill
\end{document}